\documentclass{ifacconf}

\usepackage{graphicx}      
\usepackage{natbib}        

\usepackage{float} 
\usepackage{amsmath} 
\usepackage{amssymb}  
\usepackage{mathtools} 
\usepackage{nicefrac} 
\usepackage{tikz}

\newtheorem{definition}{Definition}[section]
\newtheorem{assumption}{Assumption}[section]
\newtheorem{remark}{Remark}[section]
\newtheorem{proposition}{Proposition}[section]

\begin{document}
\begin{frontmatter}

\title{Distributed Model Predictive Safety Certification for Learning-based Control\thanksref{footnoteinfo}} 

\thanks[footnoteinfo]{This work was supported by the Swiss National Science Foundation under grant no. PP00P2\textunderscore157601/1. The research of Andrea Carron was supported
by the Swiss National Centre of Competence in Research NCCR Digital Fabrication (Agreement \#51NF40\textunderscore141853).}

\author[First]{Simon Muntwiler} 
\author[First]{Kim P. Wabersich} 
\author[First]{Andrea Carron}
\author[First]{Melanie N. Zeilinger}

\address[First]{Institute for Dynamic Systems and
	Control, ETH Z\"urich, Z\"urich, Switzerland (e-mail: \{simonmu,wkim,carrona,mzeilinger\}@ethz.ch).}

\begin{tikzpicture}[overlay, remember picture]
	\node[anchor=south,yshift=57pt] at (current page.south) {\fbox{\parbox{0.98\textwidth}{\footnotesize \textcopyright \space 2020 the Authors. This work has been accepted to IFAC for publication under a Creative Commons Licence CC-BY-NC-ND. \\
	\textbf{Cite as:} Muntwiler, S., Wabersich, K.P., Carron, A., and Zeilinger, M.N. (2020). Distributed Model Predictive Safety Certification for Learning-based Control. \textit{IFAC-PapersOnLine}, 53(2), 5258-5265. 21th IFAC World Congress. DOI: 10.1016/j.ifacol.2020.12.1205.}}};
\end{tikzpicture}

\begin{abstract}                
While distributed algorithms provide advantages for the control of complex large-scale systems by requiring a lower local computational load and less local memory, it is a challenging task to design high-performance distributed control policies. Learning-based control algorithms offer promising opportunities to address this challenge, but generally cannot guarantee safety in terms of state and input constraint satisfaction. A recently proposed safety framework for centralized linear systems ensures safety by matching the learning-based input online with the initial input of a model predictive control law capable of driving the system to a terminal set known to be safe. We extend this idea to derive a distributed model predictive safety certification (DMPSC) scheme, which is able to ensure state and input constraint satisfaction when applying any learning-based control algorithm to an uncertain distributed linear system with dynamic couplings. The scheme is based on a distributed tube-based model predictive control (MPC) concept, where subsystems negotiate local tube sizes among neighbors in order to mitigate restrictiveness of the safety approach. In addition, we present a technique for generating a structured ellipsoidal robust positive invariant tube. In numerical simulations, we show that the safety framework ensures constraint satisfaction for an initially unsafe control policy and allows to improve overall control performance compared to robust distributed MPC.
\end{abstract}


\end{frontmatter}

\section{INTRODUCTION}

One of the key mechanisms in order to improve efficiency in many control applications is the coordination with other systems. Examples include power grids, production processes, or traffic systems. Centralized control of such large-scale complex systems is often computationally infeasible or undesired from a communication point of view \citep{Langbort2004}. Distributed control algorithms address this issue by exploiting the distributed structure of the system using local control inputs that only directly depend on information from neighboring subsystems. In order to achieve optimality of a control algorithm with respect to a given cost metric under such information constraints, learning-based control offers a promising approach. However, learning-based controllers generally fail to guarantee safety in terms of state and input constraint satisfaction, because optimizing the control objective can lead to critical system states or damage to the environment \citep{Amodei2016}, which limits their application to safety-critical real-world systems. Considering the example of power grid control, violations of voltage or current constraints can lead to a reduced lifetime of the grid and are commonly restricted by legislation \citep{Bhattarai2017}.

Recent research has shown advances in the area of learning-based control of centralized systems, as for example teaching a helicopter model to fly aerobatic maneuvers in \cite{Abbeel2010} or a quadrupedal robot to perform advanced locomotion tasks in \cite{Hwangbo2019}. The application of centralized learning-based control to multi-agent systems suffers from the curse of dimensionality, since the complexity of the problem scales exponentially with the number of agents \citep{Busoniu2010}. In this context, distributed learning-based controllers have been investigated, as for example in \cite{Sartoretti2019} for the application to a collective construction task. However, such advances can only be transferred to the control of safety-critical systems once constraint satisfaction can be guaranteed. Safe learning in centralized control systems was intensively studied, e.g., in \cite{Fisac2019}, \cite{Berkenkamp2017}, but generally does not fit into the information structure of distributed control.

This paper introduces a distributed model predictive safety certification (DMPSC) mechanism for distributed linear systems with couplings in their dynamics and bounded additive disturbances, where the disturbances are essential in order to compensate for uncertainties in the system description. The proposed framework certifies safety of any distributed learning-based control input online by continuously updating a backup trajectory leading towards a set known to be safe, where the first element of the input trajectory equals the current learning-based control input. The trajectory computation is based on distributed model predictive control (MPC) concepts. If no backup trajectory exists, the learning-based input needs to be altered in order to keep the system safe. The MPC safety approach allows for unifying the computation of the backup trajectory and the required modification \citep{Wabersich2018}. 

\textit{Contributions:} The proposed safety framework combines the idea of safety certification through an MPC backup trajectory introduced in \cite{Wabersich2018} with robust distributed MPC according to \cite{Conte2013}. The distributed MPC scheme computes a nominal state trajectory and applies local feedback to keep the system state within a structured ellipsoidal robust positive invariant (RPI) tube around this nominal trajectory. Considering such a robust MPC approach renders the proposed safety framework capable of handling uncertain model descriptions, which is common in the context of learning, where system models are often derived from data. In order to reduce conservatism when verifying a local learning-based input as being safe to apply, we introduce a negotiation procedure allowing to distribute the size of the local ellipsoidal sets among neighboring subsystems. Further, we describe an adapted synthesis procedure of an RPI tube, which allows to simultaneously compute a structured feedback control law and optimize the shape of the tube.

\textit{Related Work:} Learning-based control algorithms have been specifically adapted to ensure safety by adding safety constraints in the optimization or adapting the exploration process, see e.g. \cite{Garcia2015} for a survey. A more general notion of safety is incorporated by safety frameworks as introduced in \cite{Seto1998}, consisting of a set of safe system states and a known safe control law able to keep the system within this set. The model predictive safety certification (MPSC) scheme introduced in \cite{Wabersich2018}, on which we build in this work, belongs to this category. It scales well in the state dimension and, while the concept is related to the frameworks in \cite{Akametalu2014}, \cite{Wabersich2018-2}, \cite{Fisac2019}, it allows to enlarge any given safe set \citep{Wabersich2018}.

A distributed safety framework based on structured ellipsoidal safe sets was introduced in \cite{Larsen2017}. While the approach is less computationally demanding than the proposed method, the ellipsoidal structure of the safe sets can render the certification more conservative, depending on the shape of the state constraints. The distributed safety framework proposed in this work is especially beneficial if the optimal operation point of a system is near to the border of the state constraints.

\textit{Structure:} We start by introducing the problem and the idea of the MPSC framework in Section \ref{sec:problem_description}. In Section \ref{sec:DMPSC}, we derive our distributed safety framework. Section \ref{sec:distributed_tube_design} shows the modified computation method for a distributed ellipsoidal RPI tube. Section \ref{sec:examples} gives the results of our numerical simulations and  \ref{sec:conclusion} concludes the paper.

\section{PROBLEM STATEMENT}
\label{sec:problem_description}

\subsection{Notation}

The set of integers in the interval $\left[a,\infty\right) \subseteq \mathbb{N}$ is denoted as $\mathcal{I}_{\ge a}$. The cardinality of a set $\mathcal{I} \subseteq \mathbb{N}$ is denoted as $|\mathcal{I}|$. A stacked vector $v \in \mathbb{R}^n$ consisting of subvectors $v_i \in \mathbb{R}^{n_i}$ with $i \in \mathcal{I} \subseteq \mathbb{N}$ is denoted as $v = {\text{col}}_{i \in \mathcal{I}}(v_i)$, a block-diagonal matrix $M \in \mathbb{R}^{n \times n}$ consisting of blocks $M_i \in \mathbb{R}^{n_i \times n_i}$ with $i \in \mathcal{I} \subseteq \mathbb{N}$ as $M = \mathrm{diag}_{i \in \mathcal{I}}(M_i)$. The Minkowski sum of two sets $\mathcal{A}_1, \mathcal{A}_2 \subseteq \mathbb{R}^n$ is denoted as $\mathcal{A}_1 \oplus \mathcal{A}_2 = \{a_1 + a_2 \in \mathbb{R}^n| a_1 \in \mathcal{A}_1, a_2 \in \mathcal{A}_2 \}$ and the Pontryagin difference as $\mathcal{A}_1 \ominus \mathcal{A}_2 = \{x \in \mathbb{R}^n| x + a_2 \in \mathcal{A}_1, \forall a_2 \in \mathcal{A}_2 \}$. Given a set $\mathcal{A} \subseteq \mathbb{R}^n$ and matrix $M \in \mathbb{R}^{l\times n}$ we define $M\mathcal{A} = \{x \in \mathbb{R}^l | x = Ma, a \in \mathcal{A} \}$.

\subsection{Distributed Linear Systems}

We consider a network of $M \in \mathbb{N}$ time-invariant coupled linear subsystems with discrete-time dynamics
\begin{equation}
x_i(t+1) = \left(\sum_{j=1}^{M}A_{ij}x_j(t)\right) + B_i u_i(t) + G_i w_i(t),
\label{eq:local_system_dynamics}
\end{equation}
where $x_i(t) \in \mathbb{R}^{n_i}$, $u_i(t) \in \mathbb{R}^{m_i}$ and $w_i(t) \in \mathbb{R}^{p_i}$ are the state, input and bounded disturbance of subsystem $i$ at time step $t$, respectively, $A_{ij} \in \mathbb{R}^{n_i \times n_j}$, $B_i \in \mathbb{R}^{n_i \times m_i}$, and $G_i \in \mathbb{R}^{n_i \times p_i}$. We denote the set of indices of all subsystems as $\mathcal{M}= \{1, \ldots, M\}$. The set of neighbors $\mathcal{N}_i$ of subsystem $i$ contains index $i$ itself, as well as  all indices of subsystems $j$, for which $A_{ij}$ includes entries not equal to zero. The local system dynamics of subsystem $i$ can be written as
\begin{equation}
x_i(t+1) = A_{\mathcal{N}_i}x_{\mathcal{N}_i}(t) + B_i u_i(t) + G_i w_i(t),
\label{eq:neighborhood_system_dynamics}
\end{equation}
where $A_{\mathcal{N}_i} \in \mathbb{R}^{n_i \times n_{\mathcal{N}_i}}$ and $x_{\mathcal{N}_i}(t) = \mathrm{col}_{j \in \mathcal{N}_i}(x_j(t)) \in \mathbb{R}^{n_{\mathcal{N}_i}}$. Each subsystem $i$ is subject to polytopic state and input constraints
\begin{subequations}
	\begin{align}
	\mathcal{X}_{\mathcal{N}_i} & = \{x_{\mathcal{N}_i} \in \mathbb{R}^{n_{\mathcal{N}_i}}|H_{\mathcal{N}_i}x_{\mathcal{N}_i} \le h_{\mathcal{N}_i}\} \subseteq \mathbb{R}^{n_{\mathcal{N}_i}},\\
	\mathcal{U}_i & = \{u_i \in \mathbb{R}^{m_i}|O_iu_i \le o_i \} \subseteq \mathbb{R}^{m_i},
	\end{align}
\end{subequations}
where $H_{\mathcal{N}_i} \in \mathbb{R}^{q_{\mathcal{N}_i} \times n_{\mathcal{N}_i}}$, $h_{\mathcal{N}_i} \in \mathbb{R}^{q_{\mathcal{N}_i}}$, $O_i \in \mathbb{R}^{r_i \times m_i}$ and $o_i \in \mathbb{R}^{r_i}$. Both state and input constraints contain the origin in their interior. The local disturbances $w_i(t)$ are contained in the ellipsoidal sets $\mathcal{W}_i = \{w_i \in \mathbb{R}^{p_i} | w_i^{\top}Q_iw_i \le q_i \}$ with $Q_i \in \mathbb{R}^{p_i \times p_i}$ and $q_i \in \mathbb{R}$. The global dynamics of the network is described by
\begin{equation}
x(t+1) = Ax(t) + Bu(t) + Gw(t),
\label{eq:central_system_dynamics}
\end{equation}
where $A \in \mathbb{R}^{n \times n}$, $B = \mathrm{diag}_{i \in \mathcal{M}}(B_i) \in \mathbb{R}^{n \times m}$, $G = \mathrm{diag}_{i \in \mathcal{M}}(G_i) \in \mathbb{R}^{n \times p}$, $x(t) = \mathrm{col}_{i \in \mathcal{M}}(x_i(t)) \in \mathbb{R}^n$ is the global state, $u(t) = \mathrm{col}_{i \in \mathcal{M}}(u_i(t)) \in \mathbb{R}^m$ the global input and $w(t) = \mathrm{col}_{i \in \mathcal{M}}(w_i(t)) \in \mathbb{R}^p$ the global bounded additive disturbance. The global state, input and disturbance are subject to the convex compact constraint sets
\begin{subequations}
	\begin{align}
	\mathcal{X} & = \{x \in \mathbb{R}^n|Hx \le h\} ,
	\label{eq:dist_constraints_state} \\
	\mathcal{U} & = \mathcal{U}_1 \times \ldots \times \mathcal{U}_M = \{u \in \mathbb{R}^m | O u \le o \} ,
	\label{eq:dist_constraints_input} \\
	\mathcal{W} & = \mathcal{W}_1 \times \ldots \times \mathcal{W}_M ,
	\label{eq:dist_constraints_disturbance}
	\end{align}
	\label{eq:global_constraints}%
\end{subequations}
where $H \in \mathbb{R}^{q \times n}$, $h \in \mathbb{R}^{q}$, $O = \mathrm{diag}_{i \in \mathcal{M}}\left(O_i\right) \in \mathbb{R}^{r \times m}$ and $o = \mathrm{col}_{i \in \mathcal{M}}\left(o_i\right) \in \mathbb{R}^{r}$. We assume the system matrices ($A, B$) to be stabilizable and the system state to be fully measurable.

\begin{remark}
	Considering system models with additive disturbances allows to handle systems with external disturbances and systems with uncertainty in the system parameters. It also allows to simplify even perfectly known system models, e.g., by reducing the size of the state space or neglecting nonlinearities, in order to reduce the computational load of solving the problem in a distributed fashion. Furthermore, it would allow for inexact primal dual optimization according to \cite{Koehler2019}.
\end{remark}

\subsection{Communication and Distributed Optimization}

Bidirectional communication among subsystems $i$ and $j$ is possible if $i \in \mathcal{N}_j$ or $j \in \mathcal{N}_i$. We further assume the underlying communication graph to be connected, meaning there is a communication path between any two subsystems. In order to apply the DMPSC scheme in a distributed manner, we require a distributed optimization algorithm which only uses direct communication between neighboring subsystems. An overview of parallel and distributed computation is given in \cite{Bertsekas1989}.

\subsection{Model Predictive Safety Certification}

In the following, we briefly review the ideas behind the MPSC scheme introduced in \cite{Wabersich2018}, which serves as the basis for the presented distributed safety framework. The MPSC scheme guarantees safety in terms of state and input constraint satisfaction for a linear system (\ref{eq:central_system_dynamics}) with constraints (\ref{eq:global_constraints}). The idea is to find a safe set of states $\mathcal{S} \subseteq \mathcal{X}$, for which we know a safe backup control law $u_{\mathcal{B}}$ for every state in the set. This backup control law is able to keep the system state within $\mathcal{S}$ and therefore ensures constraint satisfaction. A learning-based input $u_{\mathcal{L}}$ is applied to the system if it keeps the system state within $\mathcal{S}$. Otherwise, we can use the safe backup control law $u_{\mathcal{B}}$. The safe set $\mathcal{S}$ and safety control law $u_\mathcal{S}$ are formally introduced in the following definition.
\begin{definition}
	A set $\mathcal{S} \subseteq \mathcal{X}$ is called a safe set for system (\ref{eq:central_system_dynamics}) if there exists a known safe backup control law $u_\mathcal{B}: \mathbb{R}^n \times \mathbb{R}^m \times \mathcal{I}_{\ge 0} \rightarrow \mathcal{U}$ such that for an arbitrary policy $u_\mathcal{L}: \mathcal{I}_{\ge 0} \rightarrow \mathbb{R}^m$, the application of the safety control law
	\begin{equation*}
	u_\mathcal{S}(t) \coloneqq
	\begin{cases}
	u_\mathcal{L}(t), \text{ if } u_\mathcal{L} \in \mathcal{U} \wedge \{Ax + Bu_\mathcal{L}\} \oplus G\mathcal{W} \subseteq \mathcal{S}  \\
	u_\mathcal{B}\left(x(t), u_\mathcal{L}(t), t\right), \text{ otherwise }
	\end{cases}
	\end{equation*}
	guarantees that the system state $x(t)$ is contained in $\mathcal{S} \subseteq \mathcal{X}$ for all $t \ge \bar{t}$ if $x(\bar{t}) \in  \mathcal{S}$.
	\label{def:safe_set}
\end{definition}

In the MPSC scheme, the safe set $\mathcal{S}$ is implicitly described as the set of states where we know an MPC trajectory leading the system to a safe terminal set $\mathcal{S}_f \subseteq \mathbb{R}^n$, where constraint satisfaction can be guaranteed for all times. Robust MPC according to \cite{Mayne2005} is used, in order to account for uncertainty in the system model. The scheme can be formulated as one MPSC optimization problem, which, for a given system state $x(t)$ and learning-based input $u_{\mathcal{L}}(t)$, optimizes over an MPC trajectory leading the system to the terminal set. The objective of the problem is to minimize the `distance' between the learning-based input and the safety control law $u_\mathcal{S}(t)$. If possible, $u_\mathcal{S}(t)$ is chosen equal to $u_{\mathcal{L}}(t)$, otherwise a minimal safety-ensuring modification is computed.

The goal in this paper is to ensure constraint satisfaction of an uncertain distributed linear system (\ref{eq:local_system_dynamics}) while applying a distributed control law $u_{\mathcal{L}_i}$, for which we extend the MPSC scheme to distributed systems in the following. In this paper, we assume $u_{\mathcal{L}_i}$ to be learning-based. In principle, $u_{\mathcal{L}_i}$ could represent any control policy, e.g., human control inputs. 

\section{Distributed Model Predictive Safety Certification}
\label{sec:DMPSC}

The distribution of the MPSC scheme is based on robust distributed MPC as introduced in \cite{Conte2013}. Among the main challenges in order to transfer the centralized MPSC scheme to the distributed case are ensuring recursive feasibility and obtaining an RPI set in order to account for disturbances. In this section, we will first introduce the robust distributed MPC backup control law. Then we state the distributed safety certification problem and introduce the proposed negotiation procedure among subsystems and the form of the terminal constraint. We conclude the section by giving the theoretical safety guarantee of our distributed framework.

\subsection{Robust Distributed Backup Control Law}
The robust distributed MPC backup control law aims at keeping the distributed system (\ref{eq:local_system_dynamics}) near to the nominal system
\begin{equation}
z_i(k + 1|t) = A_{\mathcal{N}_i} z_{\mathcal{N}_i}(k|t) + B_i v_i(k|t) \quad \forall \quad k \in \mathcal{I}_{\ge 0},
\label{eq:nominal_neighborhood_system_dynamics}
\end{equation}
where $z_i(k|t) \in \mathbb{R}^{n_i}$, $z_{\mathcal{N}_i}(k|t) = \text{col}_{j \in \mathcal{N}_i}(z_j(k|t)) \in \mathbb{R}^{n_{\mathcal{N}_i}}$ and $v_i(k|t) \in \mathbb{R}^{m_i}$ are the nominal state, neighborhood state and input of subsystem $i$ at prediction step $k$ computed at time step $t \in \mathbb{Z}$. We denote the deviation between real and nominal local system state as $e_i(k|t) = x_i(t + k) - z_i(k|t) \in \mathbb{R}^{n_i}$, and similarly $e_{\mathcal{N}_i}(k|t) \in \mathbb{R}^{n_{\mathcal{N}_i}}$ and $e(k|t) \in \mathbb{R}^{n}$ for the neighborhood and global deviation.

\begin{assumption}
	There exists a stabilizing linear state feedback control law with distributed structure for system (\ref{eq:central_system_dynamics}) of the form $u = K_\Omega x = \mathrm{col}_{i \in \mathcal{M}}(K_{\Omega,i}x_{\mathcal{N}_i})$, where $K_\Omega \in \mathbb{R}^{m \times n}$ and $K_{\Omega,i} \in \mathbb{R}^{m_i \times n_{\mathcal{N}_i}}$.
	\label{ass:dist_feedback_law}
\end{assumption}

Based on $K_{\Omega}$, the tube-based control law
\begin{equation}
\begin{split}
u&(k|t) = v(k|t) + K_{\Omega}\left(x(t+k) - z(k|t)\right) \\
&= v(k|t) + \text{col}_{i \in \mathcal{M}} \left(K_{\Omega, i}\left(x_{\mathcal{N}_i}(t+k) - z_{\mathcal{N}_i}(k|t)\right)\right)
\end{split}
\label{eq:distributed_control_law}
\end{equation}
keeps the real system state within a robust positive invariant (RPI) tube around the nominal system state.
\begin{definition}\label{def:rpi_central}
	A set $\Omega \subseteq \mathbb{R}^n$ is called a robust positively invariant (RPI) set for the error dynamics if
	\begin{equation}
	e(0|t) \in \Omega \Rightarrow e(k|t) \in \Omega
	\end{equation}
	for all $k \in \mathcal{I}_{\ge 0}$ and all disturbances $w(k) \in \mathcal{W}$.
\end{definition}
In order to obtain a distributed problem structure we consider ellipsoidal RPI sets. An ellipsoidal RPI set of the form $\Omega = \{e \in \mathbb{R}^n | e^{\top} P e \le 1 \}$, with $P \in \mathbb{R}^{n \times n}$ needs to fulfill the implication 
\begin{equation}
\begin{split}
\begin{rcases}
e(k|t)^{\top} P e(k|t) \le 1 \\
w_i(t+k)^{\top} Q_i w_i(t+k) \le q_i \forall i \in \mathcal{M}
\end{rcases} & \\
\Rightarrow e(k+1|t)^{\top} P e(k+1&|t) \le 1,
\end{split}
\label{eq:rpi_implication}
\end{equation}
which is equivalent to (19) in \cite{Conte2013}. The following defines a structured RPI set.
\begin{definition}
	An ellipsoidal RPI set $\Omega \subseteq \mathbb{R}^n$ is structured if $\Omega = \{e \in \mathbb{R}^n | \sum_{i=1}^{M} e_{\mathcal{N}_i}^{\top} P_{\mathcal{N}_i} e_{\mathcal{N}_i} \le 1 \}$ with all $P_{\mathcal{N}_i} \in \mathbb{R}^{n_{\mathcal{N}_i} \times n_{\mathcal{N}_i}}$ positive semi-definite, i.e., if the shape matrix has a distributed structure.
	\label{def:structured_RPI_set}
\end{definition}

Given such a structured RPI set $\Omega$, we can define local sets $\Omega_{\mathcal{N}_i} = \{e_{\mathcal{N}_i} \in \mathbb{R}^{n_{\mathcal{N}_i}}| e_{\mathcal{N}_i}^{\top} P_{\mathcal{N}_i}e_{\mathcal{N}_i} \le \beta_i\}$. The global condition $e(k|t) \in \Omega$ is equivalent to
\begin{equation}
\forall i \in \mathcal{M} \quad \exists \quad \beta_i \ge 0: \quad e_{\mathcal{N}_i}(k|t) \in \Omega_{\mathcal{N}_i}, \sum_{i=1}^{M} \beta_i \le 1
\label{eq:beta_condition}
\end{equation}
which follows by summing up the inequalities of the local sets $\Omega_{\mathcal{N}_i}$.

Equation (23) in \cite{Conte2013} states the synthesis problem to obtain a structured ellipsoidal RPI set according to Defnition \ref{def:structured_RPI_set}. The objective of the problem can be chosen to find an RPI set with minimal volume, provided a structured control law according to Assumption \ref{ass:dist_feedback_law} is known. A possible way to find such a control law is described in \cite{Conte2016}. In Section \ref{sec:distributed_tube_design} we provide a design method that allows for simultaneously finding a distributed feedback control law according to Assumption \ref{ass:dist_feedback_law}, which can importantly reduce conservatism.

The set $\Omega$ is used to obtain the tightened constraint sets $\bar{\mathcal{X}} = \mathcal{X} \ominus \Omega$ and $\bar{\mathcal{U}} = \mathcal{U} \ominus K_{\Omega} \Omega$ for the nominal system in order to ensure $(x,u) \in \mathcal{X} \times \mathcal{U}$ when applying (\ref{eq:distributed_control_law}). The structure of $\Omega$ can be exploited to perform the constraint tightening by distributed optimization as introduced in \cite{Conte2013}, resulting in local tightened constraint sets $\bar{\mathcal{X}}_{\mathcal{N}_i}$ and $\bar{\mathcal{U}}_i$.

\subsection{Distributed Safety Certification Problem}

Using the robust distributed MPC control law we can write the DMPSC problem
\begin{subequations}
	\begin{align}
	\min_{\textbf{z}, \textbf{v}, \tilde{u}, \tilde{\beta}, \Delta\beta} & \sum_{i=1}^{M} \| u_{\mathcal{L}_i} - \tilde{u}_i \| \\
	\text{s.t. } & \forall i \in \mathcal{M}: \nonumber \\
	& x_{\mathcal{N}_i}(t) \in z_{\mathcal{N}_i}(0|t) \oplus \Omega_{\mathcal{N}_i}(t) , \label{eq:dmpsc_1_initial_constraint}\\
	&	\sum_{j \in \mathcal{N}_i\backslash i}\frac{\Delta\beta_j}{|\mathcal{N}_j| - 1} = 0 \label{eq:dmpsc_1_sum_delta} \\
	& \tilde{\beta}_i = \beta_i(t) + \Delta\beta_i \label{eq:dmpsc_1_beta_tilde} \\
	& \tilde{\beta}_i \ge 0 \label{eq:dmpsc_1_beta_tilde_ge_0} \\
	& \forall k \in \{0, \ldots, N-1\}: \nonumber \\
	& \quad z_i(k + 1|t) = A_{\mathcal{N}_i}z_{\mathcal{N}_i}(k|t) + B_iv_i(k|t) \label{eq:dmpsc_1_dynamics} \\
	& \quad \left(z_{\mathcal{N}_i}(k|t),v_i(k|t)\right) \in \bar{\mathcal{X}}_{\mathcal{N}_i} \times \bar{\mathcal{U}}_i \label{eq:dmpsc_1_constraints} \\
	& z_i(N|t) \in \mathcal{X}_{f,i}(\alpha_i(t)) \label{eq:dmpsc_1_terminal_constraint}\\
	& \tilde{u}_i = v_i(0|t) + K_{\Omega, i}(x_{\mathcal{N}_i}(t) - z_{\mathcal{N}_i}(0|t)), \label{eq:dmpsc_1_control_law}
	\end{align}
	\label{eq:dmpsc_1}%
\end{subequations}
where $\tilde{u} = \mathrm{col}_{i \in \mathcal{M}}(\tilde{u}_i)$, $\textbf{z} = \mathrm{col}_{k \in \{0,\ldots,N\}}(z(k|t))$ and $\textbf{v} = \mathrm{col}_{k \in \{0,\ldots,N-1\}}(v(k|t))$. The objective of the optimization is to find a distributed control input $\tilde{u}_i$ as `close' as possible to a distributed learning-based input $u_{\mathcal{L}_i}$. Constraint (\ref{eq:dmpsc_1_initial_constraint}) ensures that the system state lies within a tube around the initial nominal state, (\ref{eq:dmpsc_1_dynamics}) enforces the nominal dynamics and (\ref{eq:dmpsc_1_constraints}) the tightened nominal constraints, (\ref{eq:dmpsc_1_terminal_constraint}) ensures that the final nominal state lies within the terminal set and finally (\ref{eq:dmpsc_1_control_law}) defines $\tilde{u}_i$ to be a tube-based control law. We will discuss the negotiation procedure behind constraints (\ref{eq:dmpsc_1_sum_delta})-(\ref{eq:dmpsc_1_beta_tilde_ge_0}) and the details of the terminal constraint (\ref{eq:dmpsc_1_terminal_constraint}) in the following subsections. We denote the feasible set of (\ref{eq:dmpsc_1}) by
\begin{equation}
\mathcal{X}_N \coloneqq \{x \in \mathbb{R}^n | (\ref{eq:dmpsc_1_initial_constraint}) - (\ref{eq:dmpsc_1_control_law}) \quad \forall i \in \mathcal{M}\} \subseteq \mathcal{X}.
\label{eq:dmpsc_1_safe_set}
\end{equation}
The resulting certified distributed control law for each subsystem $i$ is given by
\begin{equation}
\kappa_i(x_{\mathcal{N}_i}(t)) = \tilde{u}^*_i (x_{\mathcal{N}_i}(t)),
\label{eq:dmpsc_1_optimal_controller}
\end{equation}
where $^*$ denotes the optimal solution of the problem (\ref{eq:dmpsc_1}).

\subsection{Negotiation}
In order to ensure recursive feasibility, robust MPC according to \cite{Mayne2005} constrains the initial nominal state such that the state of the system lies in a tube around it, i.e., $x(t) \in z(0|t) \oplus \Omega$. For the distributed problem in \cite{Conte2013}, this is enforced by constraint (\ref{eq:dmpsc_1_initial_constraint}) based on local sets $\Omega_{\mathcal{N}_i}(t) = \{e_{\mathcal{N}_i}(0|t) \in \mathbb{R}^{n_{\mathcal{N}_i}}| e_{\mathcal{N}_i}(0|t)^{\top} P_{\mathcal{N}_i} e_{\mathcal{N}_i}(0|t) \le \tilde{\beta}_i(t)\}$ with $\tilde{\beta}_i(t): \mathbb{N} \rightarrow \left[0,1\right]$ and $\sum_{i=1}^{M}\tilde{\beta}_i(t) \le 1$. The main goal for the DMPSC scheme is to choose the distributed control input $\tilde{u}_i$ as `close' as possible to the learning-based input $u_{\mathcal{L}_i}$. Since $\tilde{u}_i$ depends on $z_{\mathcal{N}_i}(0|t)$, we want the constraint (\ref{eq:dmpsc_1_initial_constraint}) to be as little restrictive as possible. Ideally, we would therefore add the constraint $\sum_{i=1}^{M}\tilde{\beta}_i(t) \le 1$ to the optimization problem (\ref{eq:dmpsc_1}), which is, however, a centralized condition. In order to enable distributed optimization, we introduce $\tilde{\beta}_i(t)$ as
\begin{equation}
\tilde{\beta}_i(t) = \beta_i(t) + \Delta\beta_i
\end{equation}
where $\beta_i(t)$ is a predefined function satisfying $\sum_{i=1}^{M}\beta_i(t) \le 1$ and the value of $\Delta\beta_i$ can be negotiated with neighboring subsystems, while ensuring that $\sum_{i = 1}^{M}\Delta\beta_i = 0$ in a distributed manner. By definition, this guarantees that $\sum_{i=1}^{M}\tilde{\beta}_i(t) \le 1$. This formulation enables larger values $\tilde{\beta}_i(t)$ for subsystems, where we need more flexibility to choose $z_{\mathcal{N}_i}(0|t)$ in order to represent a learning-based input.

A candidate solution for $\beta_i(t)$ introduced in \cite{Conte2013} is
\begin{equation}
\beta_i(t) := e_{\mathcal{N}_i}^*(1|t-1)^{\top} P_{\mathcal{N}_i} e_{\mathcal{N}_i}^*(1|t-1) \quad \forall i \in \mathcal{M}
\label{eq:beta_update}
\end{equation}
where $e_{\mathcal{N}_i}^*(k|t) = x_{\mathcal{N}_i}(t + k) - z_{\mathcal{N}_i}^*(k|t)$, $\beta_i(0)>0$ and $ \sum_{i=1}^{M}\beta_i(0) \le 1$. Using Proposition 1 in \cite{Mayne2005} it is possible to show that $ \sum_{i=1}^{M}\beta_i(t) \le 1$. For the negotiation parameter $\Delta\beta_i$ we introduce the local constraint (\ref{eq:dmpsc_1_sum_delta}) following the distributed information structure.
\begin{lem}
	Let the neighborhood graph be undirected, i.e., if $i \in \mathcal{N}_j$ it follows that $j \in \mathcal{N}_i$. If the local constraints  (\ref{eq:dmpsc_1_sum_delta}) are enforced locally, then the global constraint $\sum_{i=1}^{M}\Delta\beta_i=0$ is fulfilled.
	\label{lem:sum_delta_beta}
\end{lem}

\begin{pf}
	We introduce an auxiliary matrix $M_a \in \mathbb{R}^{M \times M}$ defined as
	\begin{equation}
	[M_a]_{ij} = \begin{cases}
	\frac{\Delta\beta_j}{|\mathcal{N}_j| - 1} & \text{if } j \in \mathcal{N}_i \backslash i \\
	0 & \text{otherwise,}
	\end{cases}
	\end{equation}
	where $|\mathcal{N}_j|$ denotes the number of subsystems in the set $\mathcal{N}_j$. The matrix $M_a$ contains in each row $i$ the terms $\nicefrac{\Delta\beta_j}{|\mathcal{N}_j|-1}$ for all $j \in \mathcal{N}_i \backslash i$ and in each column $j$ only terms $\nicefrac{\Delta\beta_j}{|\mathcal{N}_j|-1}$. Because of the undirected graph structure, each column contains $|\mathcal{N}_j| - 1$ entries. If we multiply $M_a$ by a vector of ones $\textbf{1} \in \mathbb{R}^M$, we obtain a vector containing the local constraints
	\begin{equation}
	M_a \textbf{1} = \begin{bmatrix}
	\sum_{j \in \mathcal{N}_1\backslash 1}\frac{\Delta\beta_j}{|\mathcal{N}_j| - 1} \\
	\vdots \\
	\sum_{j \in \mathcal{N}_M\backslash M}\frac{\Delta\beta_j}{|\mathcal{N}_j| - 1}
	\end{bmatrix} = \begin{bmatrix}
	0 \\
	\vdots \\
	0
	\end{bmatrix},
	\end{equation}
	where the right-hand side is enforced locally. We can see that $\textbf{1}^{\top} M_a \textbf{1} = 0$. Otherwise, if we multiply the matrix $M_a$ by $\textbf{1}^{\top}$ from the left first, the $|\mathcal{N}_j| - 1$ entries in each column $j$ sum up to $\Delta\beta_j$. This results in
	\begin{equation}
	\textbf{1}^{\top} M_a \textbf{1} = \sum_{i=1}^{M}\Delta\beta_i = 0,
	\end{equation}
	which shows that enforcing the local constraints ensures the global constraint.  \qed
\end{pf}

\begin{remark}
	The local constraints provide a sufficient, but not a necessary condition. At the cost of being more conservative, (\ref{eq:dmpsc_1_sum_delta}) is suited for distributed optimization. For example, for the circular graph structure in Figure \ref{fig:graph_structures}(a) the constraint (\ref{eq:dmpsc_1_sum_delta}) enforces all $\Delta\beta_i$ to be $0$. For the centered graph structure in Figure \ref{fig:graph_structures}(b) the constraint (\ref{eq:dmpsc_1_sum_delta}) gives almost the same degree of freedom for choosing $\Delta\beta_i$ as the global constraint by only enforcing $\Delta\beta_1 = 0$ and $\sum_{i=2}^{5} \Delta\beta_i = 0$.
	
	\begin{figure}[H]
		\centering
		\includegraphics[width=0.8\columnwidth]{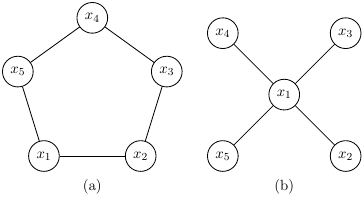}
		\caption{Examples of undirected neighborhood structures.}
		\label{fig:graph_structures}
	\end{figure}
\end{remark}

\subsection{Structured Terminal Constraint}

In robust MPC according to \cite{Mayne2005}, the terminal nominal state is constrained to an invariant terminal set, i.e., $z(N|t) \in \mathcal{X}_f$, in order to ensure feasibility and stability. Since the real system state lies within a tube around the nominal state we can define the terminal safe set as $\mathcal{S}_f = \mathcal{X}_f \oplus \Omega$.

Following from the existence of a stabilizing feedback control law for system (\ref{eq:central_system_dynamics}) according to Assumption \ref{ass:dist_feedback_law}, a structured ellipsoidal terminal invariant set $\mathcal{X}_f \subseteq \mathcal{X}\ominus \Omega$ with structured linear terminal feedback control law $K_f$ can be computed using the approach in \cite{Conte2016} based on linear matrix inequalities. This allows to write a local terminal constraint for all $i \in \mathcal{M}$ as
\begin{equation}
z_i(N|t) \in \mathcal{X}_{f,i}\left(\alpha_i(t)\right) = \left\{z_i \in \mathbb{R}^{n_i} | z_i^{\top}P_{f,i}z_i \le \alpha_i(t) \right\}
\label{eq:distributed_terminal_constraint}
\end{equation}
with the level set update rule
\begin{equation}
\alpha_i(t+1) = \alpha_i(t) + z_{\mathcal{N}_i}(t)^{\top}\Gamma_{\mathcal{N}_i}z_{\mathcal{N}_i}(t) \quad \forall i \in \mathcal{M}
\end{equation}
with $\sum_{i = 1}^{M}\alpha_i(0) \le \alpha$ and $\alpha_i(0) \ge 0$. A synthesis procedure to obtain $K_f$, $P_{f,i}$, $\alpha$ and $\Gamma_{\mathcal{N}_i}$ is given in \cite{Conte2016}.

\subsection{Safety Guarantee}
Based on the properties of distributed tube-based MPC according to \cite{Conte2013}, problem (\ref{eq:dmpsc_1}) defines a recursively feasible distributed safety framework with (\ref{eq:dmpsc_1_optimal_controller}) as safety control law.

\begin{thm}\label{thm:proof_safety}
	If Assumption \ref{ass:dist_feedback_law} holds, then (\ref{eq:dmpsc_1_optimal_controller}) is a safety control law with (\ref{eq:dmpsc_1_safe_set}) as the corresponding safe set according to Definition \ref{def:safe_set}. It also holds that (\ref{eq:dmpsc_1_safe_set}) is an RPI set.
\end{thm}

\begin{pf}
	The proof follows the line of the proof of Theorem V.1 in \cite{Conte2013} and Theorem III.7 in \cite{Wabersich2018}. Assuming (\ref{eq:dmpsc_1}) is feasible at time step $t$ with $x(t) \in \mathcal{X}_N$ we obtain an optimal solution $\textbf{z}^*$, $\textbf{v}^*$, $\tilde{u}_i^*$ and $\Delta\beta_i^*$. We start by showing that the application of $\tilde{u}_i^*$ results in $x(t+1) \in \mathcal{X}_N$, for which we can again find a feasible solution and ensure recursive feasibility using the properties of the terminal set.
	
	As shown in Lemma \ref{lem:sum_delta_beta}, constraint (\ref{eq:dmpsc_1_sum_delta}) ensures that $\sum_{i=1}^{M} \Delta\beta_i^* = 0$. Combined with $\sum_{i=1}^{M}\beta_i(t) \le 1$ and (\ref{eq:dmpsc_1_beta_tilde}) this  ensures that $\sum_{i=1}^{M}\tilde{\beta}_i(t) \le 1$. Using (\ref{eq:dmpsc_1_initial_constraint}) and condition (\ref{eq:beta_condition}) it holds that $x(t) \in z^*(0|t) \oplus \Omega$. Following Proposition 1 in \cite{Mayne2005} the application of the tube-based control law (\ref{eq:dmpsc_1_control_law}) with the nominal state dynamics (\ref{eq:dmpsc_1_dynamics}) results in $x(t+1) \in z^*(1|t)\oplus \Omega$. Because of condition (\ref{eq:beta_condition}) this can be written depending on local information as
	\begin{equation}
	\sum_{i=1}^{M} e_{\mathcal{N}_i}^*(1|t)^{\top}  P_{\mathcal{N}_i} e_{\mathcal{N}_i}^*(1|t) \le 1,
	\end{equation}
	which by the definition of the update rule (\ref{eq:beta_update}) shows that $\sum_{i = 1}^{M}\beta_i(t+1) \le 1$. The tightened nominal state constraints (\ref{eq:dmpsc_1_constraints}) ensure $x(t+1) \in \mathcal{X}$.
	
	Also by the definition of the update rule (\ref{eq:beta_update}) we know that at time step $t+1$ the initial nominal state $z(0|t+1) = z^*(1|t)$ is a feasible solution ensuring that (\ref{eq:dmpsc_1_initial_constraint}) is fulfilled by choosing the negotiation parameters $\Delta\beta_i = 0$. Using the properties of the terminal set $\mathcal{X}_f$ and control law $\sigma_f(z)$ a feasible solution at time step $t+1$ for all possible $x(t + 1)$ is the series of nominal states $\{z^*(1|t), \ldots, z^*(N|t), Az^*(N|t)+B\sigma_f(z^*(N|t))\}$, nominal inputs $\{v^*(1|t), \ldots , v^*(N-1|t), \sigma_f(z^*(N|t))\}$ and the tube-based control law $(v^*(1|t) + K_{\Omega}(x(t+1) - z^*(1|t)))$. This shows by induction that the problem (\ref{eq:dmpsc_1}) is recursively feasible and (\ref{eq:dmpsc_1_optimal_controller}) is a safety control law with (\ref{eq:dmpsc_1_safe_set}) as the corresponding robust positively invariant safe set according to Definition \ref{def:safe_set}. \qed
\end{pf}

Similarly to Corollary III.8 in \cite{Wabersich2018} it is possible to state here that by the recursive feasibility of (\ref{eq:dmpsc_1}) a longer horizon $N$ will lead to a larger or at least equally sized safe set \citep{Wabersich2018}.

\begin{remark}
	In the case of no uncertainty in the system model, the backup control law reduces to a nominal MPC scheme according to \cite{Conte2016}. Such a nominal safety framework reduces the complexity of practical implementations, while still allowing for well performing learning-based controllers for distributed systems. It also allows to handle systems with soft constraints.
\end{remark}

\section{Design of Structured Tube}
\label{sec:distributed_tube_design}

In this section we present a design procedure to obtain a structured ellipsoidal tube $\Omega$ based on \cite{Larsen2017} and \cite{Limon2008}. The key advantage of the approach is that it allows to simultaneously compute a structured feedback control law $K_{\Omega}$ and optimize the shape of the tube for minimum constraint tightening. A related method has been introduced in \cite{Koehler2017} with the difference that we consider ellipsoidal disturbance sets here. The synthesis problem introduced in \cite{Conte2013}, in contrast, assumes that a structured feedback control law according to Assumption \ref{ass:dist_feedback_law} is given. Obtaining such a control law following \cite{Conte2016} may, depending on the distributed system model, not lead to a tube fitting within the constraints even for small disturbance sets $\mathcal{W}_i$.

To simplify the notation, we define lifting matrices to switch from global to local state space, $x_i = T_i x$ with $T_i \in \{0,1\}^{n_i \times n}$, from global to neighborhood state space, $x_{\mathcal{N}_i} = V_i x$ with $V_i \in \{0,1\}^{n_{\mathcal{N}_i} \times n}$, and from global to local disturbances space, $w_i = W_i w$ with $W_i \in \{0,1\}^{p_i \times p}$.

We define a structured ellipsoidal RPI set $\Omega^{a} = \{e \in \mathbb{R}^n | \sum_{i=1}^{M} e_{i}^{\top} P_{i} e_{i} \le 1 \}$ with all $P_{i} \in \mathbb{R}^{n_{i} \times n_{i}}$ positive semi-definite, i.e., $P$ has a block-diagonal structure by definition, which can be obtained via the synthesis problem
\begin{subequations}
	\begin{align}
	\min_{E, K, h, o, \tau,  \beta^s, \beta^{s+}} \quad & h^{\top}h + o^{\top}o \label{eq:ellipsoidal_objective}\\
	\text{s.t. } & \sum_{i=1}^{M} \beta_i^s = 1, \sum_{i=1}^{M} \beta_i^{s+} = 1 \label{eq:omega_l_consensus} \\
	& \forall i \in \mathcal{M}: \nonumber \\
	& \quad \begin{bmatrix}
	\tau_{M+1} \bar{E}_i & 0 & C_i^{\top} \\
	0 & \tau_i Q_i & G_i^{\top} \\
	C_i & G_i & E_i
	\end{bmatrix} \succeq 0 \label{eq:omega_l_1_1}\\
	& \quad \tau_{M + 1}\beta_i^s - \beta_i^{s+} + \tau_i q_i \le 0 \label{eq:omega_l_1_2} \\
	& \quad \tau_i \ge 0, \tau_{M+1} \ge 0 \label{eq:omega_l_1_3}\\
	& \quad \beta_i^s \ge 0, \beta_i^{s+} \ge 0 \label{eq:omega_l_beta_i} \\
	& \quad \forall j \in \{1, \ldots, n_{h_i}\}: \nonumber \\
	& \quad \begin{bmatrix}
	\left[h_{\mathcal{N}_i}\right]_j^2 & \left[H_{\mathcal{N}_i}\right]_j E_{\mathcal{N}_i} \\
	E_{\mathcal{N}_i}\left[H_{\mathcal{N}_i}^{\top}\right]_j & E_{\mathcal{N}_i}
	\end{bmatrix} \succeq 0 \label{eq:omega_l_state} \\
	& \quad \forall j \in \{1, \ldots, n_{o_i}\}: \nonumber \\
	& \quad \begin{bmatrix}
	\left[o_i\right]_j^2 & \left[O_iK_i\right]_j \\
	\left[K_i^{\top}O_i^{\top}\right]_j & E_{\mathcal{N}_i}
	\end{bmatrix} \succeq 0 \label{eq:omega_l_2}
	\end{align}
	\label{eq:ellipsoidal_local_rpi_opt}%
\end{subequations}
where $E_i = P_i^{-1}$, $E_{\mathcal{N}_i} = P_{\mathcal{N}_i}^{-1}$, $P_{\mathcal{N}_i} = \mathrm{diag}_{j \in \mathcal{N}_i}(P_j)$, $K_i = K_{\Omega,i} E_{\mathcal{N}_i}$, $\bar{E}_i = V_i T_i^{\top} E_i T_i V_i^{\top} \in \mathbb{R}^{n_{\mathcal{N}_i} \times n_{\mathcal{N}_i}}$ is lifted to the neighborhood space and $C_i = A_{\mathcal{N}_i} E_{\mathcal{N}_i} + B_i K_i$. $\tau_i \ge 0$ and $\tau_{M + 1}\ge0$ are S-procedure parameters. The constraints (\ref{eq:omega_l_consensus})-(\ref{eq:omega_l_2}) are similar to (9b)-(9e) in \cite{Larsen2017}, where constraints (\ref{eq:omega_l_consensus})-(\ref{eq:omega_l_beta_i}) encode the RPI implication (\ref{eq:rpi_implication}) and (\ref{eq:omega_l_state})-(\ref{eq:omega_l_2}) ensure that the resulting tube lies within the polytopic state and input constraints with right-hand side $h$ and $o$. The result of the optimization problem \eqref{eq:ellipsoidal_local_rpi_opt} is the RPI set $\Omega^a$, defined by the optimizers $P_i^*$ and $K_i^*$, $i \in \mathcal{M}$. We can obtain the distributed stabilizing feedback control law $K_{\Omega,i}$ as $K_{\Omega,i} = K_i^* P_{\mathcal{N}_i}^*$.

The difference to \cite{Larsen2017} lies in the objective (\ref{eq:ellipsoidal_objective}), which is adapted from \cite{Limon2008} and in connection with the constraints (\ref{eq:omega_l_state}) and (\ref{eq:omega_l_2}) is chosen to minimize the tightening of the polytopic state and input constraints by the tube $\Omega^a$. Note that the objective \eqref{eq:ellipsoidal_objective} uses a unitary weighting under the assumption that the state and input constraints are given in a normalized form. Problem (\ref{eq:ellipsoidal_local_rpi_opt}) thereby allows to optimize over the shape of $\Omega^a$ (defined by $P$) and the structured feedback control law ($K_{\Omega,i} = K_i P_{\mathcal{N}_i}$) at the same time.

\begin{proposition}\label{prop:rpi}
	Consider $P_i^*$ and $K_i^*$, $i \in \mathcal{M}$, resulting from problem \eqref{eq:ellipsoidal_local_rpi_opt}. The structured ellipsoidal set $\Omega^a = \{e \in \mathbb{R}^n | \sum_{i=1}^{M} e_{i}^{\top} P_{i}^* e_{i} \le 1 \}$ is an RPI set for system \eqref{eq:neighborhood_system_dynamics} under the distributed linear control law $u_i(t) = K_{\Omega,i}x_{\mathcal{N}_i}(t)$ with $K_{\Omega,i} = K_i^* P_{\mathcal{N}_i}^*$.
\end{proposition}

\begin{figure*}[!t]
	\begin{equation}
	\sum_{i=1}^{M}\left(
	\begin{bmatrix}
	A_{cl}^{\top} \bar{P}_{i} A_{cl} & A_{cl}^{\top} \bar{P}_{i} G & 0 \\
	G^{\top} \bar{P}_{i}A_{cl} & G^{\top} \bar{P}_{i} G & 0 \\
	0 & 0 & -\beta_i^{s+}
	\end{bmatrix} + \tau_i \begin{bmatrix}
	0 & 0 & 0 \\
	0 & -\bar{Q}_i & 0 \\
	0 & 0 & q_i
	\end{bmatrix} + \tau_{M+1} \begin{bmatrix}
	-\bar{P}_{i} & 0 & 0 \\
	0 & 0 & 0 \\
	0 & 0 & \beta_i^s
	\end{bmatrix}\right) \text{\reflectbox{$\succeq$}} 0,
	\label{eq:K_generation_distributed_condition}
	\end{equation}
	\begin{subequations}
		\begin{align}
		\begin{bmatrix}
		\left(A_{\mathcal{N}_i} + B_iK_{\Omega,i}\right)^{\top} P_i \left(A_{\mathcal{N}_i} + B_iK_{\Omega,i}\right) - \tau_{M + 1} V_iT_i^{\top}P_iT_iV_i^{\top} & \left(A_{\mathcal{N}_i} + B_iK_{\Omega,i}\right)^{\top} P_i G_i \\
		G_i^{\top} P_i \left(A_{\mathcal{N}_i} + B_iK_{\Omega,i}\right) & G_i^{\top}P_iG_i - \tau_i Q_i
		\end{bmatrix}  \text{\reflectbox{$\succeq$}} & 0, \label{eq:larsen_cond} \\
		\tau_{M + 1}\beta_i^s - \beta_i^{s+} + \tau_i q_i \le & 0. \label{eq:larsen_cond_2}
		\end{align}
		\label{eq:dist_condition_2}
	\end{subequations}
	\hrulefill
\end{figure*}

\begin{pf}
	The ellipsoidal set $\Omega^{a}$ is an RPI set according to Definition \ref{def:rpi_central} if it fulfills \eqref{eq:rpi_implication}, which we show in the line of the proof of Lemma 4 in \cite{Larsen2017}. We introduce variables $\beta_i^s \ge 0$, $\beta_i^{s+} \ge 0$ with $\sum_{i=1}^{M} \beta_i^s = 1$, $\sum_{i=1}^{M} \beta_i^{s+} = 1$, to transform the corresponding terms in \eqref{eq:rpi_implication} to $\sum_{i=1}^{M}e_{i}(k|t)^{\top} P_{i} e_{i}(k|t) \le \sum_{i=1}^{M} \beta_i^s$ and $\sum_{i=1}^{M}e_{i}(k+1|t)^{\top} P_{i} e_{i}(k+1|t) \le \sum_{i=1}^{M} \beta_i^{s+}$. Using the S-procedure (see e.g., \cite{Boyd1994}) we can write (\ref{eq:rpi_implication}) as the linear matrix inequality in (\ref{eq:K_generation_distributed_condition}), where $\bar{P}_{i} = T_i^{\top}P_iT_i$ and $\bar{Q}_i = W_i^{\top}Q_iW_i$ are lifted to global variable spaces, and $A_{cl} = \left(A + BK_{\Omega}\right)$. Due to the sparse structure of $A_{cl}$, we can write \eqref{eq:K_generation_distributed_condition} as the two sufficient conditions in (\ref{eq:dist_condition_2}) for all $i \in \mathcal{M}$. Multiplying (\ref{eq:larsen_cond}) by $\mathrm{diag}\left(P_{\mathcal{N}_i}^{-1}, I\right)$ from both, the left and right, and applying the Schur complement (see e.g., \cite{Boyd1994}) results in constraint (\ref{eq:omega_l_1_1}). Therefore, constraints (\ref{eq:omega_l_consensus})-(\ref{eq:omega_l_beta_i}) enforce the implication \eqref{eq:rpi_implication} and ensure that $\Omega^a$ is an RPI set. \qed
\end{pf}
Based on the solution $P_i^*$ of problem \eqref{eq:ellipsoidal_local_rpi_opt}, it is possible to introduce local sets $\Omega_i^a = \{e_i \in \mathbb{R}^{n_i} | e_{i}^{\top} P_{i}^* e_{i} \le \beta_i^a \}$ such that $e(k|t) \in \Omega^a$ is equivalent to
\begin{equation}
\forall i \in \mathcal{M} \quad \exists \quad \beta_i^a \ge 0: \quad e_{i}(k|t) \in \Omega_{i}^a, \sum_{i=1}^{M} \beta_i^a \le 1.
\label{eq:beta_a_condition}
\end{equation}
Note that the existence of some local $\beta_i^a$ for a given $e(k|t) \in \Omega^a$ is not related to the variables $\beta_i^{s*}$ and $\beta_i^{s+*}$, which are introduced when designing \eqref{eq:ellipsoidal_local_rpi_opt} to derive distributed sufficient conditions for the global invariance condition \eqref{eq:rpi_implication}.
\begin{remark}
		While introducing $\beta_i^{s}$ and $\beta_i^{s+}$ in \eqref{eq:ellipsoidal_local_rpi_opt} can improve the synthesis of $\Omega^a$, it requires the global consensus constraints in \eqref{eq:omega_l_consensus}. The additional design flexibility can, e.g., be beneficial when the scale of the local disturbances differs significantly, by allowing for a growing set for systems with larger disturbances, when other systems can compensate for the growth. Depending on the problem, $\beta_i^{s}$ and $\beta_i^{s+}$ can also be fixed to fully distribute \eqref{eq:ellipsoidal_local_rpi_opt}, e.g., to $\nicefrac{1}{M}$, while maintaining its feasibility.
\end{remark}
\begin{remark}
		The block-diagonal structure of $\Omega^a$ obtained through \eqref{eq:ellipsoidal_local_rpi_opt} is more restrictive compared to the block-sparse structure of the ellipsoidal tube introduced in \cite{Conte2013}. However, since in \eqref{eq:ellipsoidal_local_rpi_opt} we simultaneously optimize over $K_{\Omega,i}$, one could also use this distributed controller to obtain a structured ellipsoidal tube using the synthesis problem (23) in \cite{Conte2013}, with a feasible solution being  $P_{\mathcal{N}_i} = V_iT_i^{\top}P_i^*T_iV_i^{\top}$ and $\bar{S}_i = 0$.
\end{remark}

\section{NUMERICAL EXAMPLES}
\label{sec:examples}

In this section we show the behavior of the introduced distributed safety framework in numerical simulations. All experiments were run using MATLAB on an Intel Core i7 2.7 GHz machine with 20 GB of RAM. We used the YALMIP toolbox \cite{Lofberg2004} with MOSEK as solver. We show that the introduced safety certificate is able to ensure constraint satisfaction and compare the regulation of a distributed system to the origin using two different unsafe nominal controllers against directly applying robust distributed MPC according to \cite{Conte2013}.

\begin{figure}
	\centering
	\includegraphics[width=\columnwidth]{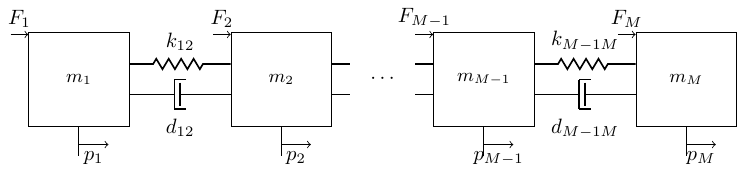}
	\caption{Chain of $M$ masses connected by springs and dampers.}
	\label{fig:mass_spring_damper_system}
\end{figure}

We consider a chain of $M$ masses $m_i$ interconnected by springs and dampers with local forces $F_i$ as inputs to the system. A schematic of the system is shown in Figure \ref{fig:mass_spring_damper_system}.

\begin{figure}
	\centering
	\includegraphics[width=0.49\columnwidth,page=1]{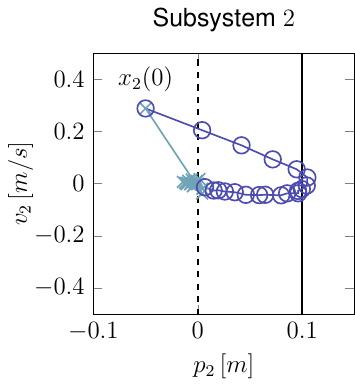}
	\includegraphics[width=0.49\columnwidth,page=2]{constraint_violation.pdf}
	\includegraphics[width=0.7\columnwidth,page=3]{constraint_violation.pdf}
	\caption{Local state space (position and velocity) of the subsystems $2$ and $8$ in the chain with system evolution over 20 time steps.}
	\label{fig:M9_spring_damper_constraint_violation}
\end{figure}

Such a system could, e.g., be used to model a vehicle platoon, which possibly reduces energy consumption and enhances safety of road traffic \citep{Li2015}. In the following we consider a numerical example with $M=9$ masses and parameter values $m_i = 1$, $k_{ij} = 0.1$ and $d_{ij} = 0.1$. The local disturbances are constrained to $\mathcal{W}_i = \{w_i \in \mathbb{R}^2 | w_i^{\top}w_i \le 1.1e-3 \}$. We use a discrete-time model obtained by Euler forward method with sampling time of $0.2 [s]$. To show constraint satisfaction we use a nominal feedback control law according to \cite{Conte2016} as input $u_{\mathcal{L}_i}$ to regulate the local subsystems to their local origins. This control law can be obtained from data by estimating the mean system  dynamics. The local positions of the subsystems are constrained to $|p_i| \le 1[m]$, with the exception of subsystem $2$, where we set $-1[m] \le p_2 \le 0.1[m]$, the local velocities to $|v_i| \le 1[m/s]$ and the inputs to $|u_i| \le 5$. The local states are relative to a local coordinate system, which potentially is moving with a constant velocity. The introduced synthesis procedure in Section \ref{sec:distributed_tube_design} with $\tau_{M + 1} = 0.055$ and setting $\beta_i^{s} = \beta_i^{s+} = \nicefrac{1}{M}$ allowed to obtain a structured ellipsoidal tube contained within the state constraints. The synthesis procedure described in \cite{Conte2013} does not allow to adapt the shape of the tube to the tight constraint of subsystem $2$, and did not result in a feasible tube to tighten the constraints for the same system parameters. We simulate the system for $20$ time steps with a prediction horizon $N=10$ for the safety framework. In Figure \ref{fig:M9_spring_damper_constraint_violation} we can see that the distributed control law $u_{\mathcal{L}_2}$ slows down subsystem $2$ very slowly and therefore leads to violation of the state constraints. The DMPSC framework is able to slow down the system on a short horizon and keep it safely around the origin. For subsystem $8$ we can see, that the safety framework does not change the evolution of the local system and allows to directly apply $u_{\mathcal{L}_8}$.

\begin{figure}
	\centering
	\includegraphics[width=\columnwidth]{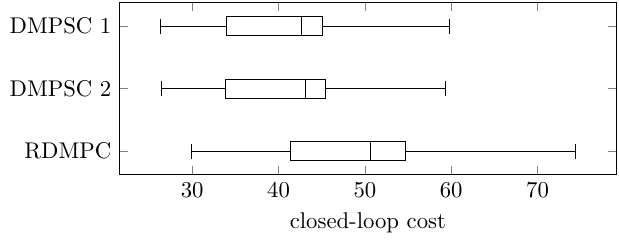}
	\caption{Box-plot of accumulated closed-loop costs.}
	\label{fig:boxplot}
\end{figure}

For the same numerical system example, we analyzed the closed-loop performance according to local stage costs $l_i(x_{\mathcal{N}_i},u_i)=\nicefrac{1}{2}\cdot x_{\mathcal{N}_i}^\top I x_{\mathcal{N}_i} + u_i^\top I u_i$ where $I$ is an identity matrix with matching dimension. We compare two different well performing policies, where the safety framework is needed to ensure constraint satisfaction, with a robust distributed MPC scheme, which is able to handle state and input constraints. The three different controller variants are: The linear feedback control law combined with the safety certificate as in the example before (DMPSC 1), a nominal distributed MPC controller without terminal cost and constraints, combined with the safety certificate (DMPSC 2) and a robust distributed MPC control law according to \cite{Conte2013} (RDMPC). We use a prediction horizon of $N=10$ for the DMPSC, the nominal and the robust MPC control law. The box-plot in Figure \ref{fig:boxplot} shows the resulting closed-loop costs for $20$ simulations with different initial positions over $20$ simulation steps. We can see that the cases DMPSC 1 and 2 have lower costs than the case RDMPC. This shows that using the safety framework in connection with an unsafe controller can lead to a better performance. For our implementation, the case DMPSC 1 on median needed 35\% less solver time compared to the case RDMPC. For the case DMPSC 2 the solver time was only about 7\% lower, since finding the nominal MPC control law is more computationally intensive than a linear feedback control law.

\section{CONCLUSION}
\label{sec:conclusion}
In this paper we introduced a distributed safety framework, which is able to ensure constraint satisfaction of distributed linear systems with coupled dynamics and bounded additive disturbances in connection with any learning-based controller. In our numerical examples we showed, that this can potentially lead to a better performance in terms of cost, while being computationally more efficient.

\bibliography{ifacconf}             

\end{document}